\documentclass{article}
\def\p{\partial}

\def\g{\gamma}

\def\De{\Delta}
\def\ld{\lambda}

\def\e{\eta}

\def\om{\omega}
\def\Om{\Omega}
\def\rh{\rho}

\def\b{\beta}

\def\a{\alpha}

\def\pdellx'{\frac{\partial}{\partial x'}}
\def\pdellw'{\frac{\partial}{\partial w'}}
\newcommand{\be}{\begin{equation}}
\newcommand{\ee}{\end{equation}}
\def\bed{\begin{displaymath}}
\def\eed{\end{displaymath}}
\def\bea{\begin{eqnarray}}
\def\eea{\end{eqncrray}}
\def\[{$$}
\def\]{$$}

\begin{document}

\title{\large \bf  Exact Rotational Space-time Transformations,\\ 
Davies-Jennison Experiments and \\ Limiting Lorentz-Poincar\'e Invariance } 
%\end{center}
\author{ Leonardo Hsu\\
Department of Postsecondary Teaching and Learning, \\
University of Minnesota, Minneapolis, Minnesota 55414 \\
and\\
 Jong-Ping Hsu\\%\footnote{E-mail: jhsu@umassd.edu}\\
 Department of Physics, \\
 University of Massachusetts Dartmouth,\\ 
North Dartmouth, MA 02747} 
%\begin{center} 

%Transformations for rotating frames.....
\bigskip
\maketitle
{\small Jennison deduced from the rotational experiments that a rotating radius $r_r$ measured by the rotating observer is contracted by $r_r = r(1-\om^2 r^2/c^2)^{1/2}$, compared with the radius $r$ measured in an inertial  frame.   This conclusion differs from the result based on Lorentz transformations.  Since rotational frames are not equivalent to inertial frames, we analyze the rotational experiments by using the exact rotational space-time transformations rather than the Lorentz transformations.  We derive exact rotational transformations on the basis of  the principle of limiting Lorentz-Poincar\'e invariance.  The exact rotational transformations form  a pseudo-group rather than the usual Lie group.  They support Jennison's contraction of a rotating radius and are consistent with two Davies-Jennison experiments.  We also suggest new experimental tests for the exact rotational transformations. }
 
\section{Introduction } 

It is evident that the space-time transformations between, say, a non-inertial and an inertial frames must simplify to the Lorentz transformations in the limit of zero accelerations.  This limiting property appears to be necessary in order  to have operational meaning for the space-time coordinates of non-inertial frames. Furthermore, it will also pave the way to formulate theories such as unified electroweak theory and gravitational theory and to understand physics in non-inertial frames.

The question is then how to generalize Lorentz transformations to non-inertial frames such as frames with linear accelerations or rotations.  To accomplish this generalization, one must find a general principle for physical laws in non-inertial frames $F(w,x,y,z)$, similar to the principle of relativity for physical laws in inertial frames $F_I(w_I,x_I,y_I,z_I)$.  In this connection, it is convenient to use $w$ with the  unit of length as the evolution variable for both inertial and non-inertial frames of reference,\cite{1} so that we can avoid the complicated properties of the speed of light (which is constant only in inertial frames) measured in terms of the usual unit, i.e., meter per second, in non-inertial frames.  

 It appears natural to postulate that all physical laws, including the laws of space-time coordinates and energy-momentum transformations, in constant-linear-acceleration frames become the corresponding laws in inertial frames in the limit of zero acceleration.  This was called the principle of limiting Lorentz-Poincar\'e invariance.\cite{2,3}  For more complicated and general non-inertial frames, we first classify reference frames into different classes.  For example, we may have the class of frames with an arbitrary-linear-acceleration (ALA) $\mbox{\boldmath$\a$}$$(w)$, the class of frames with a constant-linear-acceleration (CLA) $\mbox{\boldmath$\a$}$$_o$ and the class of inertial frames.  The general principle states that physical laws in ALA frames reduce to the corresponding laws in CLA frames in the limit $\mbox{\boldmath$\a$}$($w$)$ \   \to  \  $\mbox{\boldmath$\a$}$_o$.  We  called it the principle of limiting continuation for physical laws.\cite{4}  It includes the principle of limiting Lorentz-Poincar\'e invariance as a special case.

In the previous works,\cite{2,3,5} we applied the principle of limiting continuation to derive coordinate transformations between an inertial frame and a frame with a velocity and acceleration pointing in the same constant direction. They are consistent with known experiments.  We now use the same approach to derive coordinate transformations between an inertial frame and a frame that rotates with a constant angular velocity.  As we shall see, our results are not only consistent with the results of rotational experiments\cite{6,7,8,9} and high energy experiments involving unstable particles in a circular storage ring,\cite{10} but also support Pellegrini and Swift's analysis of the Wilson experiment,\cite{11} in which they point out that rotational transformations cannot be locally replaced by Lorentz transformations. 

While it is true that the Lorentz transformation has been used to analyze experiments involving rotational or orbital motion and that in some cases, the theoretical predictions are consistent with experimental results (for example in the case of calculating the lifetime dilation of unstable particles moving in a circular storage ring\cite{10}) the fact remains that, rigorously speaking, such applications are inappropriate. 

In this paper, we show first that the rotational space-time transformations can be obtained on the basis of  the principle of limiting Lorentz-Poincar\'e invariance and that the resultant transformations are exact because they reduce to the exact Lorentz transformations  in the limit of zero acceleration.  The fundamental metric tensors of the rotational frames are derived.  The set of exact rotational transformations of space-time forms a `pseudo-group,' which reduces to the usual Lie group only in the limit of zero acceleration.  Then we use these exact rotational transformations to discuss Jennison's contraction of a rotating radius\cite{8} and to analyze  the experimental results of Davies-Jennison.\cite{9}  Both Jennison's contraction of a rotating radius and the Davies-Jennison experiment are beyond the special  theory of relativity.  However, we show that they are consistent with the exact rotational space-time transformations based on the principle of limiting Lorentz-Poincar\'e invariance.   We also discuss new rotational experiments, which can help serve as a test of proposed rotational space-time transformations and the principle of continuation for physical laws.

\section{Exact rotational transformations with \\
limiting  Lorentz-Poincar\'e invariance } 

Suppose $F_{I}(w_{I},x_{I},y_I,z_I)$ is an inertial frame and $F(w,x,y,z)$ (which we subsequently refer to as $F(\Om)$) is a frame that rotates with a constant angular velocity $\Om$ (to be defined more precisely below). The origins of both frames coincide at all times and we use a Cartesian coordinate system in both frames.  The usual classical transformation equations between $F_I$ and $F$ are 
$$
w_I = w,  \ \ \ \ \ \  x_I = x \  cos(\Om w) - y \ sin(\Om w),$$
\be
y_I = x \ sin(\Om w) + y \ cos(\Om w),  \ \ \ \ \  z_I = z;
\ee  
%%%5.1	 
in which the relation, say, $w_I = w$ is  incorrect because it is incompatible with the experimental result for the  lifetime dilation of  unstable particle decay in flight in a circular storage ring.\cite{10}

In order to derive a set of exact coordinate transformations between $F_I$ and $F$ that satisfy the requirements of limiting Lorentz-Poincar\'e invariance (i.e., limiting four-dimensional symmetry),\footnote{This is a special case of the principle of limiting continuation of physical laws when the accelerated frame becomes an inertial frame in the limit of zero acceleration.} we first consider a slightly more general case.  In this more general case, there is an inertial reference frame $F_I$ and a non-inertial frame $F_R(\Om)$ whose origin orbits the origin of the inertial frame at a constant distance $R$, with a constant angular velocity $\Om$. A Cartesian coordinate system is used in both frames, set up in such a way that the positive portion of the $y$-axis of the $F_R(\Om)$ frame always extends through the origin of $F_I$. This is useful because in the limit $R \to \infty$ and $\Om \to 0$ such that the product $R\Om = \b_o$ is a finite non-zero constant velocity, the two frames become inertial frames with coordinates related by the familiar Lorentz transformations.  With that in mind, the classical coordinate transformations between $F_I$ and $F_R(\Om)$ (the orbiting frame) are 
$$
w_I = w,  \ \ \ \ \ \  x_I = x \  cos(\Om w) - (y-R) \ sin(\Om w),$$
\be
y_I = x \ sin(\Om w) + (y-R) \ cos(\Om w),  \ \ \ \ \  z_I = z;
\ee  
%%%5.2	 

According to the principle of limiting continuation, we postulate that the transformations between $F_I$ and $F_R(\Om)$ that satisfy the requirements of limiting Lorentz-Poincar\'e  invariance (i.e., the transformations satisfy Lorentz and Poincar\'e group properties in the limit of zero acceleration) take the following form   
$$
	w_I = Aw + B\mbox{\boldmath$\rho\cdot\b$}, \ \ \ \ \  x_I = G	x \  cos( \Om w) + E(y-R) \ sin( \Om w),$$ 	
\be
  	y_I = Ix \ sin( \Om w) + H(y-R) \ cos(\Om w),  \ \ \ \ \  z_I = z;
\ee
%%%%%%%5.3
where $\mbox{\boldmath$\rho$}=(x,y)$, ${\bf S}=(x, y-R)$, $\mbox{\boldmath$\b$}=\mbox{\boldmath$\Om$}\times  {\bf S}$, $\mbox{\boldmath$\Om$}=(0,0,\Om)$, and the functions $A, B, E, G, H,$ and $I$ may, in general, depend on the coordinates $x^\mu$. 

One unusual feature of the transformation equations (3) is that we will treat both the constant angular velocity $\Om$ (defined as $\Om=d\phi/dw$ under a suitable condition) and the orbital radius $R$ as quantities that are measured by observers in the non-inertial frame $F_R(\Om)$. This is counter to the usual procedure of measuring such parameters from the `lab' or `inertial' reference frame, but will simplify the following discussion since all quantities on the right side of the transformation equations (3) are measured with respect to $F_R(\Om)$ observers. Thus, when $w=w_I=0$, the $y$ and $y_I$ axes overlap and the origin of $F_I$, i.e., $x_I=y_I=0$, is found at the coordinates $(x, y) = (0, +R)$. Equivalently, the origin of the rotating frame $F_R(\Om)$, i.e., $x=y=0$, is found at the inertial coordinates $(x_I, y_I) = (0, -HR)$, where $H$ is an unknown function to be determined.

To determine the unknown functions $A, B, E, G, H,$ and $I$, we consider the following limiting cases. First, when $R=0$, transformations (3) must have $x/y$ symmetry (i.e., be symmetric under an exchange of $x$ and $y$). This implies that 
\be
	-E = G = H = I; 	 \ \ \ \ \ \ 	when  \ R = 0.  						
\ee
%%%%%%%5.4

Second, in the limit of small $\Om$ (or $|\mbox{\boldmath$\Om$}\times  {\bf S}| <<1$) with $R\to 0$, transformations (3) should reduce to the classical rotational transformations (1). Thus, 
\be
-E \approx G \approx H \approx I \approx 1,	 \ \ \ \ \ 	 for \ small \ \Om \ with \ R = 0.
\ee		 
%%%%%%5.5

Finally, and perhaps most importantly, in the limit where $R \to \infty$ and $\Om \to 0$ such that their product $R\Om = \b_o$ is a non-zero constant velocity parallel to the $x_I$ axis, the finite and differential forms of (3) must reduce to the Lorentz transformations, in both its finite form
\be
w_I =\g_o (w +\b_o x), \ \ \  x_I =\g_o (x +\b_o w), \ \ \  y_I = y,    \ \ \    z_I =z;
\ee
%%%%%%%%%%%%%%5.6
$$     
\g_o =\frac{1}{\sqrt{1-\b_o^2}},   
 $$ 
and its differential form, 
\be
dw_I =\g_o (dw +\b_o dx), \ \ \  dx_I =\g_o (dx +\b_o dw), \ \ \  dy_I =dy,    \ \ \    dz_I =dz.
\ee
%%5.7
(Strictly speaking, $y_I = y$ should be replaced by $y_I = -\infty$ in (6), because $R \to \infty $ in this limit.   Nevertheless, one may shift the $y$-axis so that $y_I = y$). This is necessary for (3) to satisfy the limiting Lorentz-Poincar\'e invariance, as specified by the principle of limiting continuation of physical laws.  The time $w$ in (2) reduces to the usual relativistic time (measured in the units of length) in this limit.\cite{5,1}
Thus, we have
\be
	A = B = G = -E = \g_o,	 \ \ \ \ \	H=1,   
\ee
%%%5.8						 										 
$$  when \ R \to \infty  \ and   \ \Om \to 0  \ such  \ that   \ \b_o = R  \Om.$$
 
The requirements put on $A, B, E, G, H,$ and $I$ by considering these three limiting cases do not lead to a unique solution for the unknown functions. This is analogous to the case in which gauge symmetry does not uniquely determine the electromagnetic action\cite{12} and one must also postulate a minimal electromagnetic coupling.  Here, as we did in the previous papers\cite{2,3} in deriving the accelerated Wu transformations, we postulate a minimal generalization of the classical rotational transformations (2).  

Based on the limiting cases considered above, it is not unreasonable for transformations (3) to have the following two properties: (i) For non-zero $\Om$ and finite $R$, the functions $A, B, G, I$ and $- E$ are simply generalized from $\g_o=(1-\b_o^2)^{-1/2}$ to $\g=(1-\b^2)^{-1/2}$, where $\b=S\Om$, and (ii) Only $H$ depends on $R$ and it is required to be the simplest function involving only the first power of $\g$.  Combining (4), (5), and (8) with the two properties just named leads to the following solutions   
\be
A = B = G = I = \g = (1 -\b^2)^{-1/2},   \ \ \ \ \	 E = -\g,    
\ee
%%%%%5.9
$$											
	 H = (\g+R/R_o)(1+R/R_o), \ \ \ \ \  \g=(1-\b^2)^{-1/2}.
$$
The quantity $R_o$ in (9) is an undetermined length parameter that seems to be necessary for the transformations to satisfy the limiting criteria specified in (4), (5), and (8), as well as the limiting Lorentz-Poincar\'e invariance.  From a comparison of rotational transformations and their classical approximation, the value of $R_o$ should be very large.  However, since it is unsatisfactory to have such undetermined parameters in a theory and there are no known experimental results that can help to determine a value for $R_o$, we shall take the limit $R_o \to \infty$ for simplicity.  This is consistent with the minimal generalization of the classical rotational transformations (2).  
			 
Thus, a simple rotational transformation corresponding to the classical transformations (2) is    
$$
	w_I = \g(w + \mbox{\boldmath$\rho\cdot\b$}), \ \ \ \ \  x_I = \g[x \  cos(\Om w) - (y-R) \ sin(\Om w)],$$ 	
\be
  	y_I = \g[x \ sin(\Om w) + ( y-R) \ cos(\Om w)],  \ \ \ \ \  z_I = z;
\ee
%%%%%%%5.10
$$
\b = |\mbox{\boldmath$\Om$}\times  {\bf S}|=\Om \sqrt{x^2 +(y-R)^2}=\Om S<1, \ \ \  \mbox{\boldmath$\rho\cdot\b$}=x R\Om.  
 $$
 
Since the transformations (10) include the exact Lorentz transformations as a special limiting case, as shown in (6),   the transformations (10) are also exact.  We shall call (10) `rotational taiji transformations' to distinguish it from other rotational transformations in the literature.\footnote{In ancient Chinese thought, the word `taiji' or `taichi' denotes `The Absolute,' i.e., the ultimate principle or the condition that existed before the creation of the world.} 

 Because all known experiments can be analyzed in the $R=0$ case, for the rest of this paper, we set $R=0$ and concentrate solely on the implications of the transformations under that condition. With $R=0$ the coordinate transformations between an inertial frame $F_I$ and a non-inertial frame $F(\Om)$ whose origin coincides with that of $F_I$ and that rotates with a constant angular velocity $\Om$ about the origin are
$$
	w_I = \g(w + \mbox{\boldmath$\rho\cdot\b$})=\g w,  \ \ \ \   x_I = \g[x \  cos(\Om w) - y \ sin(\Om w)],
$$	
\be
  	y_I = \g[x \ sin(\Om w) +  y \ cos(\Om w)],   \ \ \ \ \  z_I = z; \ \ \ \ \   \g = {1}/{\sqrt{1-\rh^2 \Om^2}},
\ee
%%%%%%%%11
where $\mbox{\boldmath$\rho\cdot\b$}=\Om R x =0$ for $R=0$.

To derive the inverse transformations of (11), we must first find a way to express $\Om w$ and $\g=1/\sqrt{1-\rh^2\Om^2}$ in terms of quantities measured in the inertial frame $F_I$.  While the coordinate transformations (10) for the non-zero $R$ case can only be written in Cartesian coordinates (so that they satisfy the limiting Lorentz-Poincar\'e invariance), the transformations in the $R=0$ case (11) can be written in terms of cylindrical coordinates. We introduce the relations $x_I = \rh_I \ cos\phi_I, y_I = \rh_I  \ sin\phi_I, x = \rh \ cos\phi,$ and $y = \rh \ sin\phi$ so that (11) can be written as
$$
w_I = \g w, \ \ \ \  \rh_I =\sqrt{x_I^2 +y_I^2} = \g \rh, \ \ \ \
\phi_I =\phi +\Om w, \ \ \ \   z_I = z;    
$$
with
$$
\Om_I \equiv \frac{d \phi_I}{d w_I}=\frac{d \phi_I}{d w}\frac{d w}{d w_I}=\Om \frac{1}{\g} , \ \ \ \ \ \  (\rh \ and \ \phi  \ fixed).
$$
The conditions that $\rho$ and $\phi$ are fixed mean that an object is at rest in the rotating frame, so that an observer in $F_I$ can measure its angular velocity $\Om_I$ and identify it with the angular velocity of the rotating frame. These equations give the operational definitions of $\Om_I$ and $\Om$.
Thus, we have
\be
w_I \Om_I  =w \Om , \ \ \ \   \rh_I \Om_I  =\rh \Om , \ \ \ \
\g=\frac{1}{\sqrt{1-\rh^2 \Om^2}} = \frac{1}{\sqrt{1- \rh_I^2 \Om_I^2}}.
\ee
%%%%%%%%%%%12
From (11) and (12), we can then derive the inverse rotational transformations  
$$
	w= \frac{w_I}{\g}, \ \ \ \ \  x= \frac{1}{\g}[x_I \  cos(\Om_I w_I) + y_I  \ sin(\Om_I w_I)],$$ 	
\be
  	y= \frac{1}{\g}[-x_I \ sin(\Om_I w_I) + y_I \ cos(\Om_I w_I)],  \ \ \ \ \  z= z_I;
\ee
%%%%%%%5.13
where
$$
\g=\frac{1}{\sqrt{1- \Om_I^2(x^2_I +y^2_I)}}.   
$$

If one imagines the frame $F(\Om)$ as a carousel, then for any given angular velocity, objects at rest relative to $F(\Om)$ that are sufficiently far from the origin would have a classical linear velocity that exceeds the speed of light in the inertial frame $F_I$. As expected, the rotational transformations display some unusual behavior near that region. From the relationship between $\Om$ and $\Om_I$, i.e., $\Om_I  =\Om/\g = \Om \sqrt{1-\rh^2 \Om^2}$, one can see that for a given constant angular velocity $\Om$, the corresponding angular velocity $\Om_I$ depends on $\rh$ as well as $\Om$.  In particular, when $\rh =1/\Om$,  the value of $\Om_I$ becomes zero. Furthermore, the reading of $F(\Om)$ clocks at that radius, as viewed by $F_I$ observers, stops changing, much like the case in special relativity where the rate of ticking of clocks slows down and stops as the speed of those clocks approaches the speed of light.  These effects are not unexpected because the rotational transformations (11) map only a portion of the space in the rotating frame $F(\Om)$ ($\rh <  1/\Om$) to the entire inertial frame $F_I$. Similar to the singular walls found in frames with a linear acceleration,\cite{2} there is a cylindrical singular wall at $\rh =1/\Om$ in the rotating frame $F(\Om)$.

The rotational transformations (11) have the inverse transformations (13), but when $R \ne 0$ in (10), it appears that there is no algebraic expression for the inverse of the transformations (10). However, mathematically, (10) defines a function from one neighborhood of the origin in space-time to another. The implicit function theorem then guarantees the existence of an inverse in a neighborhood of any point where the Jacobian is non-zero.\cite{13} 

\bigskip

\section{Metric tensors for the space-time of rotating frames}
We note that the  transformations of the contravariant 4-vectors $dx^{\mu}_I =(dw_I,dx_I,$$dy_I,dz_I)$ and $dx^{\mu}=(dw,dx,dy,dz)$ can be derived from (11). We have
$$
dw_I = \g[dw +(\g^2 \Om^2 w x)dx + (\g^2 \Om^2 w y)dy],$$
$$
dx_I =\g\left\{[cos(\Om w) + \g^2 \Om^2 x^2 cos(\Om w) - \g^2 \Om^2 xy \ sin(\Om w)]dx \right.$$
$$
-[ sin(\Om w) + \g^2 \Om^2 y^2 sin(\Om w)  - \g^2 \Om^2 xy  cos(\Om w)]dy$$
$$\left. -[\Om x \ sin(\Om w) + \Om y \ cos(\Om w)]dw\right\},$$
\be
dy_I =\g\left\{[sin(\Om w) + \g^2 \Om^2 x^2 sin(\Om w) + \g^2 \Om^2 xy \ cos(\Om w)]dx\right.
\ee
%%%%%%%5.16%%%%%%14
$$
+[ cos(\Om w) + \g^2 \Om^2 y^2 cos(\Om w)  + \g^2 \Om^2 xy \ sin(\Om w) ]dy $$ 
$$
+\left.[\Om x \ cos(\Om w) - \Om y \ sin(\Om w)]dw\right\},$$
$$
dz_I = dz. $$
To find the metric tensors $P_{\mu\nu}$, which will be called the Poincar\'e metric tensors, for the rotating frame $F(\Om)$, it is convenient to use (11) to write $ds^2=\e_{\mu\nu}dx_I^\mu dx_I^\nu= dw_I^2 - dx_I^2 - dy_I^2 - dz_I^2$,  where $\e_{\mu\nu}=(1, -1, -1, -1)$, as  
\be
ds^2 = d(\g w)^2 - (x^2+y^2)\g^2 \Om^2 dw^2 - d(\g x)^2 - d(\g y)^2 -dz^2
\ee
%%%%%5.17%%%%%%%%%%%%%%%5.15
$$
+2\g \Om y \ d(\g x) dw - 2 \Om \g x \ d(\g y) dw.$$
 Then, with the help of the relation $d\g = \g^3 \Om^2(xdx+ydy)$, (15) can be written as
 \be
ds^2 = P_{\mu\nu}dx^{\mu} dx^{\nu},	 \ \ \ \ \ \	\mu, \nu = 0, 1, 2, 3.				 
\ee
%%%%%5.18%%%%%%%%16
It follows from (14), (15) and (16) that the non-vanishing components of $P_{\mu\nu}$ are given by  
$$
P_{00} = 1,  \ \ \     P_{11} = -\g^2\left[1+2\g^2 \Om^2 x^2 - \g^4 \Om^4 x^2(w^2 -x^2 -y^2)\right],$$
$$
P_{22}= -\g^2\left[1+2\g^2 \Om^2 y^2 - \g^4 \Om^4 y^2(w^2 -x^2 -y^2)\right],\ \ \  P_{33}=-1,$$
\be
P_{01} = \g^2\left[\Om y + \g^2 \Om^2 wx], \ \ \ \ \  P_{02} =\g^2[-\Om x + \g^2 \Om^2 wy\right],
\ee
%%%%%%%5.19%%%%%%%%%%%%%t17
$$
P_{12}=-\g^4 \Om^2 xy\left[2-\g^2 \Om^2 (w^2 -x^2 -y^2)\right].$$
Using $P_{\mu\ld}P^{\ld\nu} = \delta^\nu_{\mu}$, the contravariant metric tensor $P^{\mu\nu}$ is found to be
$$
P^{00} = \g^{-2}\left[1 -\Om^4 w^2 (x^2+y^2)\right],  \ \ \ \     P^{33} = -1,$$
$$
P^{11}=  -\g^{-2}\left[\g^{-2}(1-\Om^2 x^2) -2 \g^{-2} \Om^3 wxy +\Om^6 w^2  y^2 (x^2+y^2)\right],$$
\be
 P^{22} = -\g^{-2}\left[ \g^{-2}(1-\Om^2 y^2) +2 \g^{-2} \Om^3 wxy +\Om^6 w^2  x^2 (x^2+y^2)\right], 
\ee
 %%%%5.20%%%%%%%%%%%%%18
 $$
 P^{01}= -\g^{-2}\left[-\Om y - \g^{-2} \Om^2 wx + \Om^5 w^2 y (x^2+y^2)\right],$$
 $$
  P^{02}= -\g^{-2}\left[\Om x - \g^{-2} \Om^2 wy - \Om^5 w^2 x (x^2+y^2)\right],$$
  $$
   P^{12}= \g^{-2}\left[\g^{-2} \Om^2 x y - \g^{-2} \Om^3 w(x^2 - y^2) + \Om^6 w^2 xy (x^2+y^2)\right].$$
   All other components in (17) and (18) are zero.  One can verify that these Poincar\'e metric tensors $P_{\mu\nu}$ for rotating frames satisfy vanishing Riemann-Christoffel curvature tensors.  When a rotating frame with a very large $R$ approaches an inertial frame in the limit of zero acceleration, the Poincar\'e metric tensors will reduce to the Minkowski metric tensors $\e_{\mu\nu}=(1,-1,-1,-1)$.
   
In the special case $R=0$, one can easily express the exact rotational transformations (11)  in cylindrical coordinates $x^\mu=(w, \rh, \phi, z)$ and obtain the metric tensors for the rotating frame,
$$
P_{00}=1, \ \ \  P_{11}= - \g^6(1-\Om^4\rh^2 w^2), \ \ \  P_{22}=-\g^2 \rh^2, \ \ \  P_{33}=-1,
$$
$$
P_{01}= \g^4 \Om^2 \rh w, \ \ \ \    P_{02}= -\g^2 \Om \rh^2.
$$

\section{The rotational pseudo-group}  

As mentioned previously, the rotational transformations (11) imply the existence of a cylindrical singular wall at $\rho =\sqrt{x^2+y^2}=1/\Om \equiv \rho_s$, which depends on the angular velocity $\Om$. Like the accelerated Wu transformations, the rotational transformations map only the portion of space-time of the rotating frame within the singular wall to the entire space-time of an inertial frame and the `group' of rotating transformations is a pseudo-group.\cite{14} In general, the rotational transformations map only a portion of the space in a rotating frame $F(\Om)$ to a portion of the space in another rotating frame $F(\Om')$.
To deal with these types of transformations, mathematicians O. Veblen and J. H. C. Whitehead developed the concept of a pseudo-group\cite{14}. A set of transformations forms a pseudo-group if
(i) the resultant of two transformations in the set is also in the set, (ii) the set contains the inverse of every transformation in the set.\footnote{
Here, ÒsetÓ is defined as an infinite-dimensional set, rather than the usual Lie set.} The complete set of rotational transformations forms a pseudo-group, which may be called  the rotational pseudo-group. 
%%%%%%%%%%%ref.13-14

Two of the group properties of the rotational transformations are straightforward to verify. One can easily see that in the limit $\Om \rightarrow 0$, the rotational transformations (11) reduce to the identity 
transformation.  Also, the inverse transformations of (11) are given by (13).

In order to see other group properties of the rotational transformations, let us consider two other rotating frames, $F'(\Om')$ and $F''(\Om'')$, which are 
characterized by two different constant angular velocities $\Om'$ and $\Om''$, respectively.  With the help of (11), we can derive the rotational transformations between $F(0)=F_I, F(\Om), F'(\Om')$ and $F''(\Om'')$, 
$$
w_I=\g w ={\g}' w' = {\g}'' w'', $$ 
%%%%%%%%%%
$$x_I =  \g[x \  cos(\Om w) - y \ sin(\Om w)]={\g}'[x' \  cos(\Om' w') - y' \ sin(\Om' w')]$$
\be
 =  \g''[x'' \  cos(\Om'' w'') - y'' \ sin(\Om'' w'')],
 \ee
 %%%%%%%%33%%5.19
$$
  y_I = \g[x \ sin(\Om w) + y \ cos(\Om w)]={\g}'[x' \ sin(\Om' w') + y' \ cos(\Om' w')] $$	
$$=\g''[x'' \ sin(\Om'' w'') + y'' \ cos(\Om'' w'')],$$
$$
 \g '=\frac{1}{\sqrt{1-{\b'}^2}} , \ \ \ \ \ \   {\b}' = \Om '\rho ',  \ \ \ \ \   \g ''=\frac{1}{\sqrt{1-{\b''}^2}} , \ \ \ \ \ \   {\b}'' = {\Om}''{\rho}''. 
$$
where we have neglected to write the trivial transformation equations for the $z$ direction.  The rotational transformations between $F(\Om)$ and $F'(\Om')$ can be obtained from (19):
$$
 w=\frac{\g '}{\g} w',
$$
%%%%%%%%%24%%%5.34%%%20
$$x = {\g '}\left\{[x'  \ cos(\Om ' w')-y' \sin(\Om ' w')]\frac{ cos(\Om w)}{\g}\right.$$
\be
\left. +[x' \ sin(\Om ' w') +y' \ cos(\Om ' w'))\frac{sin(\Om w)}{\g}]\right\},
 \ee
$$ y = \g ' \left\{[x'  \ sin(\Om ' w')+y' \cos(\Om ' w')]\frac{ cos(\Om w)}{\g}\right.$$
$$\left. - [x' \ cos(\Om ' w') -y' \ sin(\Om ' w')]\frac{sin(\Om w)}{\g}\right\}.$$

If one compares these transformations with those in equations (11) to (13), one can see that the transformations between two rotating frames are more algebraically complicated than those between an inertial frame and a rotating frame. In general, it does not seem possible to write the inverse of the transformations (20) in analytic form. However, the implicit function theorem guarantees the existence of an inverse in the neighborhood of any point where the Jacobian is non-zero.\cite{13} 

Moreover, the set obtained from generators of the rotational transformations using the Lie bracket, and the brackets of brackets, and so on, never closes in general. That is, the rotational transformations define a group with infinitely-many generators.\cite{13}  This difference is likely related to the distortion of space-time coordinates in non-inertial frames relative to an inertial frame, as shown by the Poincar\'e metric tensors $P_{\mu\nu}$ in (17). 

The space-time transformations for frames involving constant-linear-accelerations or arbitrary-linear-accelerations also form pseudo-groups rather than Lie groups.\cite{5}  Thus, regarding the group nature of coordinates transformations in flat space-time between general physical frames of reference, we conjecture that only in the limits of zero accelerations, we have Lie groups such as Lorentz and Poincar\'e groups associated with the space-time transformations of inertial frames.

\section{Physical implications}

Some physical implications of the rotational transformations for rotating frames worthy of note are:
\bigskip

\noindent
(i)  Operational definitions of space-time coordinates in rotating frames

Since the speed of light in rotating frames is not constant, as shown by $ds = 0$ in (15), it is very complicated to use light signals to synchronize clocks in a rotating frame $F (\Om) = F (w, x, y, z)$.  However, as discussed previously\cite{3}  in reference to linearly accelerated Wu transformations, one can synchronize a set of clocks and setup a coordinate system in the rotating frame $F(\Om)$ by using a grid of `computerized space-time clocks'\cite{4} that are programmed to accept information concerning their positions $(x_I, y_I)$ relative to the $F_I$  frame, obtain $w_I$ from the nearest $F_I$ clock,\footnote{A system of clocks in an inertial frame $F_I$ can be synchronized.} and then compute and display $w, x$ and $y$ using the inverse transformation (13) with the given parameter $\Om_I$. In a rotating frame $F(\Om)$, the values the physical time $w$ can take on are restricted by the condition $\Om^2 (x^2 + y^2) < 1$ and the coordinates of physical space in $F (\Om)$ are limited by $\rh=\sqrt{x^2+y^2} < 1/\Om$.
\bigskip

\noindent
(ii)  The invariant action for electrodynamics in rotating frames 

	We are now able to write the invariant action $S_{em}$ in natural units in a rotating frame for a charged particle with mass $m$ and charge $e$  moving in the 4-potential $A_{\mu}$: 
\be
	S_{em} = \int \left[-mds - e A_{\mu}dx^{\mu} - \frac{1}{4}F_{\mu\nu} F^{\mu\nu}\sqrt{-P}d^4 x\right],
\ee
	%%%%%%%%%%%%%%%%%5.21%%%%%%%%%31%%%%%27%%%%21
\be
	F_{\mu\nu} =D_{\mu} A_{\nu} - D_{\nu} A_{\mu}=\p_{\mu} A_{\nu} - \p_{\nu} A_{\mu}, \ \ \ \      P=det P_{\mu\nu},
\ee
%%%%%%%%%%%%%%5.22%%%%%%%%%%%%%%%32%%%%%%%%%%%28%%%22
 where $ds$ is given by (16) and the Poincar\'e metric tensors $P_{\mu\nu}$ for rotating frames are given in (17).  Also,   $D_{\nu}$ denotes the covariant partial derivative associated with the metric tensors  $P_{\mu\nu}$. 
  
 	The Lagrange equation of motion of a charged particle can be derived from (21).  We obtain
\be
m\frac{D u_{\mu}}{ds} = eF_{\mu\nu} u^{\nu}, 
\ee
\be
Du_{\mu}=D_{\nu} u_{\mu}dx^{\nu}, \ \ \ \ \   u^{\nu} = \frac{dx^{\nu}}{ds},  \ \ \ \ \   u_{\mu} = P_{\mu\nu} u^{\nu}.
\ee
%%%%%%%%%%%%%%5.24%%%%%%%%%%34%%%%%%%%%%30%%%24

	Starting with the invariant action (21) and replacing the second term $- \int e A_{\mu}dx^{\mu}$ with 
\be
	- \int A_{\mu} j^{\mu} \sqrt{-P}d^4 x
\ee
	%%%%%%%%5.25%%%%%%%%%35%%%%%%%%%%%31%%%%25
  for a continuous charge distribution in space, we obtain covariant Maxwell's equations in a rotating frame		
\be
D_{\nu} F^{\mu\nu} =  j^{\mu}, \ \ \ \ \    \p_{\ld}F_{\mu\nu}   + \p_{\mu} F_{\nu\ld}+\p_{\nu} F_{\ld\mu}=0.
\ee
%%%%%%%%%%5.26%%%%%%%%%%36%%%%%%%32%%%%%26
 	Based on gauge invariance and the rotational invariance of the action (21), the term $e A_{\mu}dx^{\mu}$ implies that the electromagnetic potential $A_\mu$ must be a covariant vector   in non-inertial frames because the coordinate differential $dx^\mu$ is, by definition,  a contravariant vector.  Since the force ${\bf F}$ and the fields ${\bf E}$ and ${\bf B}$ are related to a change of the potential $A_\mu$   with respect to a change of coordinates  $x^{\mu}$, by definition, the electromagnetic fields ${\bf E}$ and ${\bf B}$ are naturally identified with components of the covariant tensor $F_{\mu\nu}$ as given by (22) in non-inertial frames.  The metric tensor $P_{\mu\nu}$ behaves like a constant under covariant differentiation, $D_{\a}P_{\mu\nu} = 0$.  
	
	Similarly, one can also formulate classical Yang-Mills gravity in rotating frames and other non-inertial frames.\cite{15,16}  Nevertheless, the quantization of fields and the derivation of Feynman rules for QED and Yang-Mills gravity in rotating frames are difficult due to the complicated metric tensors in (17) and (18).\cite{17}  
	
\bigskip

\noindent
(iii)  Absolute contraction of a rotating radius and absolute slow-down of a rotating clock

In contrast to the classical rotational transformations (1), the exact rotational  transformations (11) predict that the length of a rotating radius $\sqrt{x^2+y^2}$ is contracted by a factor of $\g$,  
\be
\sqrt{x_I^2+y_I^2}= \g \sqrt{x^2+y^2}.
\ee
%%%%%%%%%5.15%%%%%%%%%%%%%%%%%%%%%%%%38%%34%%33%%27
This contraction is absolute, meaning that both observers in the inertial frame $F_I$ and in the rotating frame $F(\Om)$ agree that the radius, as measured in the rotating frame $F(\Om)$ is shorter, because there is no relativity between an inertial frame and a rotating (non-inertial) frame.  

In some discussions of phenomena involving circular motion at high speeds, such as the lifetime dilation of unstable particles traveling in a circular storage ring, the argument is made that during a very short time interval, one can approximate the true rotational transformations using the Lorentz transformations.\cite{10,11}  However, making this approximation leads to a completely different conclusion regarding the radius, namely, that it does not contract because it is always perpendicular to the direction of motion.

Furthermore, for a given $\rh$, (11) implies $\De w_I = \g \De w$, which is independent of the spatial distance between two events.\footnote{For comparison with the usual time t, if one were to define $ w = ct$ and $w_I = ct_I$, then one would have the usual relation $\De t_I = \g \De t$.  However, the constant `speed of light' $c$ in a non-inertial frame is not well-defined because the speed of a light signal is no longer isotropic or constant.}  In other words, clocks at rest relative to a rotating frame and located at a distance $\rho = \sqrt{x^2 + y^2}$  from the center of rotation slow down by a factor of $\g=\sqrt{ (1- \b^2)}$ in comparison with clocks in the inertial frame $F_I$.  Analogous to the absolute contraction of radial lengths as shown in (27), this time dilation is also an absolute effect in that observers in both $F_I$ and $F(\Om)$ agree that it is the accelerated clocks that are slowed. 
 
Both the contraction of radial distances and the slowing down of clocks are consequences of requiring the rotational transformations to satisfy the limiting Lorentz-Poincar\'e invariance.

It is interesting to note that,  in 1964, Jennison\cite{8} showed by deductions from the results of an experiment by Champeney and Moon that ``the radius of a rotating system, measured from a single domain rotating with the system, would be $r_r = r(l -\om^2 r^2)^{1/2}$ and that the angular velocity would be $\om_r = \om(1- \om^2  r^2 )^{-1/2}$ where $\om$ and $r$ are the angular velocity and radius in the non-rotating laboratory frame,''  (where $c=1$ in natural units).\cite{6,8} Such a Jennison contraction of rotating radius and the increase of angular velocity are consistent with our results in (12) with the angular velocity $\Om_I = \om$, $\Om=\om_r$ and the  time relations $w_I =\g w$ for the absolute slow down of a rotating clock.
\bigskip

\noindent
(iv)  The transformations of covariant momenta or wave vectors 	

	Although the speed of light is not a universal constant in a rotating frame $F(\Om)$, one can still formulate a covariant momentum $p_{\mu} = (p_0, p_i),$ $ i=1,2,3$ using the time $w$ as the evolution variable in the Lagrangian formalism.  From the invariant `free'  action $S_f = -\int mds = \int L \ dw$ for a `non-interacting particle' with mass $m$ in the rotating frame $F(\Om)$, the spatial components of the physical momentum are
\be 
p_i = -\frac{\p L}{\p v^i }=m P_{\nu i} \frac{dx^{\nu}}{ds} = P_{\nu i} p^{\nu},  \ \ \ \ \  L=-m\sqrt{P_{\mu\nu} v^\mu v^\nu}, 	
 \ee
 %%%%%%%%%5.27%%%%%%%%%%%%%%39%%%%%%%%%35%%34%%%28
 where $i=1,2,3$.  Both $L$ and $p_i$  have the dimension of mass and $v^\mu \equiv dx^\mu/dw = (1,v^i)$.  The zeroth component $p_0$ (or the Hamiltonian) with the dimension of mass is defined as usual
 \be
 p_0 = v^i\frac{\p L}{\p v^i}
- L = mP_{\nu 0} \frac{dx^\nu}{ds} = P_{\nu 0} p^{\nu}.
\ee
%%%%%%%%%%%%%%%%5.28%%%%%%%%%%%40%%%%%%%%%36%%35%%%29
 Thus, the covariant momentum $p_\mu$ defined in (28) and (29) is the physical momentum of a particle in the rotating frame.
 
	The rotational transformations of the differential operators $\p/\p x_I^{\mu}$ and  $\p/\p x^{\mu}$ can be calculated from (13).  Just as in quantum mechanics, the covariant momentum $p_{\mu}$ has the same transformation properties as the covariant differential operator $\p/\p x^{\mu}$. From (13), we obtain   
$$
p_{I0} = \g^{-1} (p_0 + \Om y p_1 -\Om x p_2),
$$
$$
p_{I1} =\left[-\g^{-2} \Om^2 wx_I\right]   p_0 + \g^{-2}\left[\g cos(\Om w) - \Om^2 x_I x - \Om^3 wx_I y\right]p_{1}
$$
%%%%%%%%29%%%%%%%%%%%%41%%%%%%%%%%%37%%36%%%%30
\be
+ \g^{-2}\left[  -\g sin(\Om w) - \Om^2 x_I y + \Om^{3} wx_I x\right] p_2,
\ee
$$
p_{I2} =\left[-\g^{-2} \Om^2 wy_I\right]   p_0 + \g^{-2}\left[\g sin(\Om w) - \Om^2 y_I x - \Om^3 wy_I y\right]p_{1}
$$
$$
+ \g^{-2}\left[ \g cos(\Om w) - \Om^2 y_I y + \Om^{3} w y_I x\right] p_2,$$
$$
p_{I3}=p_3, \ \ \ \ \   \g=\frac{1}{\sqrt{1-\rh^2 \Om^2}},
$$
where $x_I$ and $y_I$ can be expressed in terms of $x, y$, and $w$ of the rotating frame $F(\Om)$ using (11).
 
 	Consider the case of a particle traveling at high speed in a circular storage ring.  Such a particle can be considered to be at rest in a rotating frame $F(\Om)$, so that $dx^i = 0$ and hence, $ds = dw$.  Based on $p^{\nu} = mdx^{\nu}/ds$ in (28), the contravariant momenta are  $p^i = 0,  \ i = 1,2,3$, and $p^0 = m$.  In this case, the covariant momenta of this particle in $F$ are
\be
	p_0 = m, \ \ \ \ \ p_1 =m\g^2 \Om y, \ \ \ \ \ p_2 = -m\g^2 \Om x,\ \ \ \ \  p_3 = 0,
\ee
%%%%%%%%%%%%%30%%%%%%%%%%%%%42%%%%%38%%37%%%%31
where we have used   the relations $p_\mu=P_{\mu\nu}p^\nu, \   dx^i = 0, \  \rh^i=(x,y,0)=fixed,  \  \rh^2 = constant$.  Since both $\rh$ and $\g$ are constant in the rotational transformations (11) and (13), equation (14) will be greatly simplified and, hence, (17), (18), and (30) will be simplified too.  As a result, we also have very simple relations in (31).
 This difference between $p_\mu$ and $p^\mu$ is mainly due to the non-vanishing metric tensor components $P_{01}$ and $P_{02}$ in the rotating frame.  The physical momenta of the particle, as measured in an inertial frame $F_I$, are given by (30) and (31),
  $$
	p_{I0} = \g m, \ \ \ \ \ \ \ \  p_{I1} =m\g [\Om x \ sin(\Om w) +  \Om y \ cos(\Om w)],
$$
\be
p_{I2} =- m\g[\Om x \ cos(\Om w) - \Om y \ sin (\Om w)], \ \ \  p_{I3} = 0.
\ee
%%%%%%%%%%%%%31 %%%%%%%%%%%%43%%%39%%38%%%%32 
Thus, we have seen that the expression for the energy of a rotating particle $p_{I0} = \g m$ agrees with the well-established results of high energy experiments performed in an inertial laboratory frame $F_I$. 

The transformation laws of the covariant wave vector $k_\mu$ are also given by (30) because a photon's momentum $p_\mu$ and wave vector $k_\mu$ is related by $p_\mu=k_\mu$ (in natural units). 

\section{Experimental tests of the rotational taiji transformations}

Because particles moving in a straight line at relativistic speeds travel large distances in very short times, designing experiments to test relativistic effects with linear motion, where the Lorentz transformations are most directly applicable, can be challenging.  Experiments in which objects move in a circular path at relativistic speeds can be performed with much more compact apparatus. However, as was discussed briefly in the previous section, applying the Lorentz transformations to analyzing such experiments is problematic because of the non-relativity between inertial frames and rotating frames, as shown in the transformations (10), (11) and (13). In this section, we discuss three experiments that can test the usefulness of the  rotational taiji transformations (10) or (11).

\bigskip

\noindent
(A)  Absolute dilation of decay-length for particle decay in circular motion

	Let us first consider the lifetime dilation of unstable particles traveling in a circular storage ring with a constant radius.\cite{10}  If the particle's rest lifetime in the rotating frame $F(\Om)=F(w,x,y,z)$ is denoted by $\De w(rest)$, then the rotational taiji transformation (11) gives
\be
\De w_I = \g \De w(rest), \ \ \ \ \    \g=\frac{1}{\sqrt{1-\b^2}} , \ \ \ \ \ \   \b = \Om\rho. 
\ee
%%%%%%%%%%%32%%%%%%%%%%%%%%44%%%%%%40%%39%%%%33
But the rest lifetime of the particle $\De w(rest)$ in the rotating frame $F$ cannot be directly measured in the inertial laboratory.  However, according to the weak equivalence of non-inertial frames (see equation (A2) in the Appendix),\cite{4,2} we have the relaiton,   
\be
\De w(rest) \sqrt{P_{00}} = \De w_I(rest) \sqrt{\e_{00}}.
\ee
%%%45%%%%%%%%%41%%40%%%%34
 Since the lifetime is typically very short, $\De w(rest)$ can be considered as the zeroth component of the contravariant differential coordinate vector $dx^0$:
$
\De w = dw =dx^0.
$
It follows from (33) and (34) that
\be
\De w_I=\g \De w_I(rest),
\ee
%%%%%5.45%%%%%42%%41%%%%%35
where we have used $P_{\mu\nu} $ in (17).
 Result (35) is consistent with well-established experimental results of the decay lifetime (or decay length)  dilation of muons in flight in a circular storage ring.\cite{10,5} 
\bigskip

\noindent
(B)  Davies-Jennison experiment 1--null transverse frequency shift

An experiment that can test the usefulness of the taiji rotational transformations is the Davies-Jennison experiment 1,\cite{8,18} in which the shift in frequency of transversely-emitted radiation from an orbiting source was measured.  The arrangement of the apparatus for this experiment is roughly as follows:
A laser beam is directed straight downward onto a mirror at the center of a horizontal rotating table. The mirror sends the beam radially outward, parallel to the surface of the table, where it strikes a second mirror mounted at the edge of the table. This second mirror reflects the beam in a direction perpendicular to its velocity, thus acting as a source emitting radiation in the transverse direction. This resultant beam is then mixed with light from the original laser in order to produce interference fringes and measure any shift in its frequency compared to the original laser light. In the Davies-Jennison experiment 1, no such shift was observed. Let us analyze this experiment using the rotational taiji transformations.
 
Consider a radiation source at rest in a rotating frame $F(\Om)$, located at ${\mbox{\boldmath$\rho$}}_s=(x_s,y_s)$, that emits a wave with a frequency $k_0=k_0(rest)$, as measured in $F(\Om)$.  If this wave, travels along the $-{\mbox{\boldmath$\rho$}}_s$ direction to the origin, then $k_1/k_2 =k_x/k_y=x/y,$  because the two vectors $-{\mbox{\boldmath$\rho$}}$ and ${\bf k}$ are on the x-y plane and parallel to each other.  Because the rotational taiji transformations for the covariant wave vector $k_\mu$ are the same as those for the covariant momentum $p_\mu=\hbar k_\mu$, the frequency shift of transversely-emitted radiation as measured by observers in an inertial laboratory frame $F_I$ is given by (30).  To detect the transverse effect, the observer is assumed to be at rest at the origin $x_I=y_I=x=y=0$.  Equation (30) with $p_\mu=\hbar k_\mu$ gives the relation for frequencies,
\be
k_{I0}=\g^{-1}(k_0 + \Om[ y k_1 - x k_2]) = \g^{-1} k_0,
\ee
%%%10%%%12%%%36 
$$
  x k_2 = y k_1, \ \ \ \   \g=\frac{1}{\sqrt{1-\rh_s^2 \Om^2}}, \ \ \  \rho_s^2=x_s^2 + y_s^2=constant,
$$
%%%42%%%%%
where $k_0=k_0(rest)$ is the frequency of the source at rest in the rotating frame $F(\Om)$, as measured by observers in $F(\Om)$.  

Just as in our discussion of the lifetime dilation of unstable particles traveling in a circle, it is extremely difficult to measure the frequency $k_0(rest)$ directly because we cannot move our measuring apparatus to the rotating frame $F(\Om)$. Furthermore, because of the non-equivalence of inertial and rotating frames, we cannot assume that $k_0(rest)$ is equal to $k_{I0}(rest)$, the frequency of the radiation from the same source at rest in an inertial frame $F_I$ as measured by observers in $F_I$. However, we can again use the weak equivalence of non-inertial frames (cf. equation (A2) in  the Appendix)\cite{4,16} to write 

\be
k_0(rest) \sqrt{P^{00}} = k_{I0}(rest) \sqrt{\e^{00}}, \ \ \ \   P^{00}=\g^{-2},  \ \ \  \e^{00}=1.
\ee
%%%43%%%%11%%%13555%%%%%%37
Recall that $k_0(rest)$ is the frequency of radiation emitted from the source at rest in the rotating frame $F(\Om)$, located at the constant radius, $\rh=fixed$ on the x-y plane.  When this condition is used in the derivation of (14), then we have $dw_I = \g dw$, etc. so that the contravariant component $P^{00}$ in (18) reduces to $P^{00} =\g^{-2}=constant$ in (37).\footnote{For details, see equations in (39) below.}  
  It follows from (36) and (37), that the relationship between the frequency of the transversely-emitted radiation emitted from the orbiting source as measured in the inertial laboratory frame $k_{I0}$ and the frequency of the radiation emitted from the same source at rest in the inertial laboratory frame as measured in the inertial laboratory frame $k_{I0}(rest)$ is given by
\be
k_{I0}=\g^{-1}k_0(rest) =  k_{I0}(rest).
\ee
%44%%12%%%%%%%%%14%%%%5.38
Thus, the rotational taiji transformation predicts the result (38), which implies that no frequency shift for transverse radiation should be observed for an orbiting radiation source, as measured by observers in inertial frames.  This result (38) is consistent with Davies-Jennison experiment 1 \cite{8,18}. 

 \bigskip

\noindent
(C) Longitudinal frequency shifts and Davies-Jennison experiment 2

Another experiment that can test the taiji rotational transformations is the longitudinal frequency shift of an orbiting source as observed in inertial laboratory. Consider a radiation source at rest in a rotating frame $F(\Om)$, located at ${\rh^i}=(x,y, 0)=constant$, that emits a wave with a frequency $k_0=k_0(rest)$, as measured in $F(\Om)$.  Let us consider  the rotational transformations with the condition $\rh^2  = x^2+y^2=constant$, one can show that the rotational transformations for the coordinate differentials $dx^\mu$ and  the results for the metric tensors $P_{\mu\nu}$ and $P^{\mu\nu}$ in (17) and (18) will be simplified.  For our discussions, it suffices to note that $P^{\mu\nu}$ in (18) becomes
\be
P^{00}= \g^{-2}, \ \ \  P^{11}=-\g^{-2}(1-\Om^2 y^2), \ \ \  P^{22}=-\g^{-2}(1-\Om^2 x^2), 
\ee
%%%%%%%%%%%42%%%%%39
$$
 P^{01}=\g^{-2} \Om y,  \ \ \  P^{02}=-\g^{-2} \Om x,  \ \ \    P^{12}=\g^{-4} \Om^2 xy, 
 $$
 for constant $\rh$.
Similarly, the transformations of wave vector $k_\mu (= p_\mu/\hbar)$ will be simpler than (30):
$$
k_{I0} = \g^{-1} (k_0 + \Om y k_1 -\Om x k_2),
$$
$$
k_{I1} =\g^{-1}\left( k_1\left[ cos(\Om w) + \Om^4 x y \rh^2 sin(\Om w) \right]\right.
$$
%%%%%%%%29%%%%%%%%%%%%41%%%%%%%%%%%37%%36%%%%30$$$$40
\be
- \left.k_2 \left[ sin(\Om w) + \Om^4 x y  \rh^2 cos(\Om w) \right]\right) ,
\ee
$$
k_{I2} =\g^{-1}\left( k_1 \left[ sin(\Om w) - \Om^4 y x\rh^2 cos(\Om w)\right] \right.
$$
$$
+\left. k_2\left[ cos(\Om w) - \Om^4 x y\rh^2 sin(\Om w)\right]\right).
$$

We now concentrate  on the beams in the $\pm x_I$ directions (i.e., $k_{I2} = k_{I3}=0$), as observed in the laboratory.  In this case, we also have $k_2 = k_3 =0. $  For the light beam emitted at time $w=0$ from the position $(x,y) = (0, y)$, the transformation (40) leads to the longitudinal frequency shift,
\be
k_{I0}=\g^{-1} (k_0 + \Om y k_1), 
\ee
%%41
where $k_0=k_0(rest)$ is the frequency of the source at rest in the rotating frame $F(\Om)$, as measured by observers in $F(\Om)$.  

Because of the non-equivalence of inertial and rotating frames, we cannot assume that $k_0(rest)$ is equal to $k_{I0}(rest)$, the frequency of the radiation from the same source at rest in an inertial frame $F_I$ as measured by observers in $F_I$. However, we can again use the weak equivalence of non-inertial frames (37) to obtain the relation, 
\be
k_0(rest) \g^{-1} = k_{I0}(rest).    
\ee
%%%43%%%%11%%%13555%%%%%%37%%%%46%42
The law of propagation of light in $F(\Om)$, $k_\mu k_\nu P^{\mu\nu}=0$, with the source at $(x,y)=(0,y)$ with $\rh^2=x^2+y^2=constant$, leads to
\be 
k_0^2 - k_1^2 \g^{-2} + 2\Om y k_0 k_1 = 0, \ \ \ \ \   k_2=k_3=0,
\ee
%%%%%%%%%%%%%%%%%43
where we have used the metric tensor $P^{\mu\nu}$  in (39). 

 From equation (43), we solve for $k_1>0$,
\be
k_1= \g^2 k_0 (1 + \Om y), \ \ \ \ \  \g^2=\frac{1}{1-\Om^2 \rh^2},  \ \ \  \rh^2=y^2.
\ee
%%%%%%%%%%43%%%%44
 It follows from (41), (42) and (44) that the exact rotational space-time transformations predict the orbiting frequency shift to be
\be
k_{I0} = k_{I0}(rest)\left[ 1 +y \Om \g^2 (1 + \Om y)\right]
\ee
$$
\approx  k_{I0}(rest)\left[ 1 + \rh_{I} \Om_I + \rh_{I}^2\Om_I^2 ....\right], \ \ \ \   \rh\Om = \rh_{I}\Om_I,
$$
%%%%%%47%%%%%%45
where we have used $k_0 =  k_0(rest) $ for source at rest in $F(\Om)$ and  express $y \Om$ in terms of quantities measured in inertial laboratory frame $F_I$, $y \Om = y_{I}\Om_I$, according to (12) with $\rh=y$ and $\rh_I=y_I$.
Thus, the exact rotational taiji transformations predict a new frequency shift (45) for orbiting source emitting light in the longitudinal direction, as measured by observers in inertial frames.  The result (45) is in sharp contrast to the usual relativistic Doppler shift in special relativity for radiation source moving with {\em constant velocities}.  

 The arrangement of the apparatus for the Davies-Jennison experiment 2 is similar to their  experiment 1 discussed in (B), section 6.   A small modification is that the second mirror (mounted at the edge of the rotating table) reflects the beam in a direction parallel to its velocity, thus acting as a source emitting radiation in the longitudinal direction.  The Davies-Jennison experiment 2 does not measure the frequency shift (45) directly \cite{18,9}.  Rather, the experiment sends the reference beam of the interferometer directly to the detector while the other beam is reflected in turn from a mirror moving towards the beam and then from a mirror moving away from the beam, where both beams are parallel to the $x_I$.  Suppose the mirror moving away from the beam is at $(x,y)=(0,-R_s)$ and the mirror moving towards the beam is at $(x,y)=(0, +R_s)$.    The frequency of this longitudinal beam is then  compared to the reference beam in the laboratory.  

The shifts of frequencies at these two locations of orbiting mirrors are respectively given by (41) with $(x,y)=(0, -R_s)$ and $(x,y)=(0, +R_s)$:
\be
k_{I0}=\g^{-1} (k_0 + \Om y k_1)= k_{I0}(rest)\left(1+ \Om y \frac{k_1}{k_0}\right), \ \ \ \ \  y=-R_s,
\ee
%%46
\be
k_{I0}=\g^{-1} (k_0 + \Om y k_1)= k_{I0}(rest)\left(1+ \Om y \frac{k_1}{k_0}\right), \ \ \ \ \ \  y=+R_s.
\ee
%%%%%%47
Here, the ratio $k_1/k_0$ is determined by the law $k_\mu k_\nu P^{\mu\nu}=0$ for light propagation in the rotating frame, where $P^{\mu\nu}$ is given by (39).\footnote{If the cylindrical coordinates is used for the rotating frame $F(\Om)$, the law for light propagation, $k_\mu k_\nu P^{\mu\nu}=0$, implies explicitly that the ratio $k_1/k_0$ depends on the constant radius $\rh>0$ in this case (where $k_1$ is understood as proportional to the tangential wave vector).} Since $k_1$ is given by (44), the results in (46) and (47) show that the blue-shift is completely cancelled by the red-shift, as observed in the inertial laboratory frame $F_I$, so that the rotational transformations predict null frequency shift, $k_{I0}= k_{I0}(rest)$, consistent with the null result of the Davies-Jennison experiment 2 \cite{18}.  The main physical reason for this null  frequency shift is the weak equivalence of non-inertial frames, as expressed by the relation (42).

Therefore, rather than serving as a test of special relativity, the rotational Davies-Jennison experiments can reveal new principles of physics for non-inertial frames, which are beyond the realm of special relativity.\cite{19}  They can also help to develop a new and deeper understanding of physics in non-inertial frames based on the  principle of limiting  Lorentz-Poincar\'e invariance and the weak equivalence of non-inertial frames.

The work was supported in part by JingShin Research Fund of  the UMass Dartmouth Foundation.  The authors would like to thank D. Fine for useful discussions and S. H. Kim for using symbolic computing to double check (17),  (18) and (39).
\bigskip

\bigskip

\noindent
{\bf \large Appendix. \ \ \ Weak equivalence of non-inertial frames}
\bigskip

Suppose a  particle is  at rest in an inertial frame $F'_I$ and has a lifetime $\tau'_I (rest) $, as measured by observers in   $F'_I$.   But from the viewpoint of observers in another inertial (laboratory) frame $F_I$, this particle will decay in flight and has a different lifetime $\tau_I $, as measured by observers in $F_I$. 
  The Lorentz transformations  give the usual lifetime dilation $\tau_I = \g_o \tau'_I (rest).$
Experimentally, it is difficult to measure $\tau'_I(rest)$ because it is almost impossible to arrange for an observer to be in a frame that is co-moving with unstable particles in high-energy laboratory.  Fortunately, according to the principle of relativity, all inertial frames are equivalent.  Therefore, the lifetime  $\tau_I(rest)$ of a particle at rest in $F_I$ and measured by observers in $F_I$ is  the same as the lifetime $\tau'_I(rest)$ of the same kind of particle at rest in $F'_I$ and measured by observers in $F'_I$, 
$$
  \ \ \ \ \ \ \ \ \ \ \ \ \ \ \ \  \ \ \ \ \ \ \ \ \ \ \ \ \ \ \ \ \ \   \tau_I(rest)=\tau'_I(rest).   \ \ \ \ \ \ \ \ \ \ \ \ \ \ \ \  \ \ \ \ \ \ \ \ \  \ \ \ \ \ \ \  (A1)
$$
   Thus, the lifetime dilation (3.34) in special relativity can be expressed in terms of observable quantities in an inertial frame $F_I$, $ \tau_I=\g_o \tau_I(rest), $
which has been confirmed by high-energy experiments in the laboratory frame $F_I$.

Now, we consider the situation involving non-inertial frames $F$ and $F'$, which are not equivalent to an inertial frame $F_I$.  However, in order to calculate the lifetime of an accelerated particle to compare it with experiment, we must generalize the relation such as $ \tau_I(rest)=\tau'_I(rest)$ in (A1) for inertial frames to a more general relation for non-inertial frames.  

Let us consider observable quantities which transform as vectors.  These physically observable vector quantities (e.g., the energy-momentum of a particle with a mass, or the wave vector of the radiation emitted from a source, etc.) can be denoted by contravariant or covariant vectors, $V^\mu(rest)$ and $V_\mu(rest)$, where $(rest)$ denotes that the particle or the source of radiation is `at rest'  in a non-inertial frame $F$ and these quantities are measured by observers in the same non-inertial frame $F$. Following the principle of limiting continuation for physical laws, it is natural to postulate the following generalized relations for two different non-inertial frames, $F$ and $F'$,
\bed
 \ \ \ \ \ \ \ \ \ \ \ \ \ \ \ \ \ \ \ \ \ \ \  V^0(rest)\sqrt{P_{00}}=V'^0(rest)\sqrt{P'_{00}} \ ,   \ \ \ \ \ \ \ \ \ \ \ \ \ \ \ \  \ \ \ \ \ \ \ \ \  (A2)
\eed
for contravariant vector such as differential coordinate vectors, or 
$$
 \ \ \ \ \ \ \ \ \ \ \ \ \ \ \ \ \ \ \ \ \ \ \  V_0(rest)\sqrt{P^{00}} = V'_0(rest)\sqrt{P'^{00}} ,  \ \ \  \ \ \ \ \ \ \ \ \ \ \ \ \ \ \ \ \ \ \ \ \ \ \ (A3)
$$
for covariant vectors such as momentum and wave vectors.   These postulates may be termed the `weak equivalence of non-inertial frames.'  These relationships (A2) and (A3)  cannot be derived from space-time transformations for non-inertial frames because they refer to two different vectors (or physical quantities) which are not connected by the space-time coordinate transformations.  However, they are consistent with the principle of limiting continuation of physical laws and can be tested experimentally. 

 It appears that the Davies-Jennison experiments 1 and 2 discussed previously in section 6 and Thim's rotational experiment\cite{20} support the weak equivalence of non-inertial frames, as postulated in (A2) and (A3).
 We may remark  that for the special case, where $P_{\mu\nu}=0$ for $\mu \ne \nu$, (A2) is  the same as (A3).  This property of the metric tensor $P_{\mu\nu}$ shows up if the non-inertial frames are constant-linear-acceleration frames, whose space-time  transformations are given by the accelerated Wu transformations.\cite{4,5}  Also, (A2) and (A3) are consistent with (A1) in the limit of zero acceleration, where all Poincar\'e metric tensors $P_{\mu\nu}$ for non-inertial frames reduce to the Minkowski metric tensor, $\e_{\mu\nu}=(1,-1,-1,-1)$ for inertial frames.

 \newpage
%\begin{thebibliography} 
\bibliographystyle{unsrt}

\begin{thebibliography}{99}
\bibitem{13}We would like to thank Dana Fine for helpful discussions.
\end{thebibliography}

\end{document}